# Deformation of an Elastic Beam due to Viscous Flow in an Embedded Parallel Channel Network

Yoav Matia & Amir D. Gat

Faculty of Mechanical Engineering, Technion - Israel Institute of Technology
Technion City, Haifa, Israel 32000

Elastic deformation due to embedded fluidic networks is currently studied in the context of soft-actuators and soft-robotic applications. In this work, we analyze interaction between the elastic deflection of a slender beam and viscous flow within a long serpentine channel, embedded in the elastic beam. The channel is positioned asymmetrically with regard to the midplane of the beam, and thus pressure within the channel creates a local moment deforming the beam. We focus on creeping flows and small deflections of the elastic beam and obtain, in leading order, a fourth-order partial integro-differential equation governing the time-dependent deflection field. This relation enables the design of complex time-dependent deformation patterns of beams with embedded channel networks, including inertia-like standing and moving wave solutions in configurations with negligible inertia.

## 1. Problem Formulation

We consider the dynamics of an elastic beam, initially at rest, with a serpentine channel network embedded within it asymmetrically with regard to the neutral-plane. The embedded channel network is filled with a viscous fluid (Fig.1a) and time-varying pressure is introduced at the channel inlets at one or both ends. Due to the pressure-field applied by the fluid at solid-fluid interface, the channel cross section deforms and with it the external beam structure (Fig.1b).

We define beam height $h_s$, width $b_s$ and length $l_s$, and require $h_s/b_s \ll 1$ and $b_s/l_s \ll 1$ (Fig. 1a and 1b). The Young's modulus, Poisson's ratio and mass per unit length of the beam are $E$, $\nu$ and $\rho_s$, respectively. An interconnected parallel Embedded Fluidic Network (denoted EFN hereafter) is located within the beam perpendicular to $x_s - y_s$ plane (Fig. 1a). The length of a single serpentine segment is denoted $l_c$ and the width of the beam $b_s$, where $l_c/b_s \sim 1$. We limit our analysis to configurations with $l/l_c \gg 1$ and $(l_c \cdot n)/l \sim 1$, where $l$, is the total length of the serpentine channel and $n$ is the number of channel segments of length $l_c$, in order to allow relating of the discrete problem with a continuous function. The embedded channel network is assumed to have a negligible effect on the beam second moment inertia $I$ and mass per unit length. The total deflection of the beam in the $y_s$ direction is denoted by $d_s$. Assuming small deflections, we can define $d_s = d_e + d_c$, where $d_e$, is the deflections due to external forces, and $d_c$, the deflection due to the embedded channel network.

The embedded channel coordinate system $(x_c, y_c, z_c)$, is defined such that, $\hat{x}_c$-direction is the channel stream-wise direction, $h_c$ and $b_c$ denote channel characteristic height in the $\hat{y}_c$-direction and width in the $\hat{z}_c$-direction, respectively. We assume $b_c \sim h_c$ and define a small parameter representing channel slenderness $\varepsilon_1 = h_c/l \ll 1$. The physical properties of the fluid are viscosity, $\mu$, velocity, $\mathbf{u} = (u, v, w)$ and gauge pressure, $p$. Channel cross-section area is defined by the sum $a(x_c, p) = a_0(x_c) + a_1(x_c, p)$, where $a_0(x_c)$ is the cross-section area of the channel at gauge pressure $p = 0$, and $a_1(x_c, p)$ describes the change of the cross section area due to the fluid pressure.

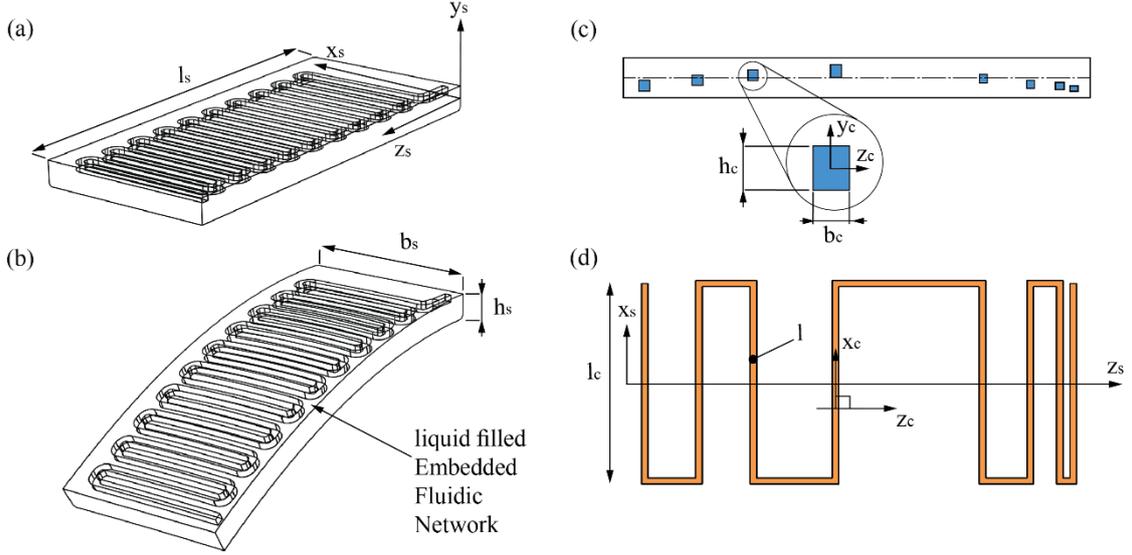

FIG. 1. An illustration of a Embedded Fluidic Network (EFN) configuration, consisting of an interconnected parallel channel network positioned asymmetrically with respect to the neutral plane. (a) EFN at rest. (b) The deflection due to uniform pressure EFN. (c) $y_s - z_s$ plane and definition of channel cross section (d) $x_s - z_s$ plane illustrating network with a varying channel density distribution.

The governing equations for incompressible creeping Newtonian flow are the Stokes equation,

$$\nabla p = \mu \nabla^2 u \qquad (1.1)$$

and conservation of mass

$$\nabla \cdot u = 0. \qquad (1.2)$$

A previous work[12], presented a modified Euler-Bernoulli equation governing the deformation field of an EFN elastic beam,

$$\frac{\partial^2}{\partial z_s^2}\left[EI\left(\frac{\partial^2 d_s}{\partial z_s^2} + \phi(z_s)\frac{p(z_s,t)}{E}\frac{\partial \psi}{\partial (p/E)}\right)\right] = -\rho_s \frac{\partial^2 d_s}{\partial t^2} + f(z_s,t). \qquad (1.3)$$

The value of $\psi$, representing the change in beam slope due to a single channel and its derivative $\partial \psi/\partial(p/E)$, have been defined in [12] and are can be obtained experimentally or via numerical computations. Equation (1.3) details a one-way coupling of fluid pressure to beam deflection, thus limiting our analysis to configurations in which the fluid pressure created by external forces acting on the beam is negligible compared with the pressure applied at the channel inlets.

Hereafter, we define normalized variables by capital letters and characteristic values by asterisk superscripts. We define the characteristic velocity, $(u^*, v^*, w^*)$, characteristic gauge pressure, $p^*$, characteristic channel cross-section at gage pressure $a_0^*$, characteristic change in channel cross-section due to characteristic pressure, $a_1^*$, viscous-elastic time scale, $t_f^*$, characteristic channel density $\phi^*$, characteristic total beam deflection, $d_s^*$, characteristic elastic-inertial time scale $t_s^*$ and characteristic external force applied to the beam, $f^*$.

We now turn to define normalized variables. Channel network spatial coordinates are denoted $(X_c, Y_c, Z_c) = (x_c/l, y_c/h_c, z_c/h_c)$, time $T = t/t_f^*$, fluid velocity (in the $(X_c, Y_c, Z_c)$ coordinates) $(U, V, W) = (u/u^*, v/v^*, w/v^*)$, pressure $P = p/p^*$, volume flow rate at channel cross-sectional $Q = q/(h^2 u^*)$, volume flow-rate at a given cross section for $\partial P/\partial X_c = 1$ is denoted $Q_1(A(X_c, P))$, channel cross section, $a(x_c, p) = a_0(x_c) + a_1(x_c, p)$ is normalized via $a_0^*$, and $a_1^*$, such that it reads $A(X_c, P) = A_0(X_c) + \sigma_r \cdot A_1(P)$ where $\sigma_r = a_1^*/a_0^* = a_1^*/h^2$, channel density $\Phi(Z_s) = \phi(z_s)/\phi^*$, beam spatial coordinates are $(X_s, Y_s, Z_s) = (x_s/b_s, y_s/h_s, z_s/l_s)$, beam total deflection $D_s = d_s/d_s^*$, external forces applied to the beam $F = f/f^*$.

We now focus our analysis on the parallel serpentine channel configuration embedded within a beam as illustrated in Fig.1c and Fig.1d. Assuming $\phi(z_s) \gg 1$, we convert the channel cooredinate $x_c$ to the $z_s$ beam coordinate as,

$$x_c(z_s) = l_c \cdot \int_0^{z_s} \phi(z_s)\, dz_s + z_s. \tag{1.4}$$

Substituting the normalized parameters, we obtain

$$X_c = \int_0^{Z_s} \Phi(Z_s)\, dZ_s + \frac{l_s}{l} \cdot Z_s, \tag{1.5}$$

and the resulting characteristic channel density is $\phi^* = l/l_c l_s$.

# 2. Governing equation of EFN beam deformation due to internal viscous flow

Substituting the normalized parameters into (1.2) and (1.1) yields in leading order,

$$\frac{\partial U}{\partial X_c} + \frac{\partial V}{\partial Y_c} + \frac{\partial W}{\partial Z_c} \sim 0, \tag{2.1}$$

$$\frac{\partial P}{\partial X_c} \sim \frac{\partial^2 U}{\partial Y_c^2} + \frac{\partial^2 U}{\partial Z_c^2}, \quad \frac{\partial P}{\partial Y_c} \sim 0, \quad \frac{\partial P}{\partial Z_c} \sim 0 \tag{2.2}$$

where $h/l \sim v^*/u^* = \varepsilon_1 \ll 1$, $u^* = p^*\varepsilon_1^2 l/\mu$. Integrating (2.1) over the channel cross section in the $Y_c - Z_c$ plane and applying Gauss theorem yields,

$$\frac{\partial Q}{\partial X_c} + \frac{h}{t_f^* v^*} \frac{\partial A}{\partial T} = 0. \tag{2.3}$$

We define $Q_1(X, A(X_c, P))$ as the solution for the Poisson equation, (2.2), using channel boundary conditions of no slip at wall $(U, V, W)|_{\text{wall}} = \underline{V}_{\text{wall}}$,

$$Q = -\frac{\partial P}{\partial X_c} \cdot Q_1(A(X_c, P)). \tag{2.4}$$

$Q_1(A(X_c, P))$ can be readily calculated for a given cross-section. From $A(X_c, P) = A_0(X_c) + \sigma_r \cdot A_1(P)$

$$\frac{\partial A}{\partial T} = \sigma_r \cdot \frac{\partial A_1(P)}{\partial P} \cdot \frac{\partial P}{\partial T}, \tag{2.5}$$

Substituting (2.5) and (2.4) into (2.3),

$$-\left(\frac{\partial^2 P}{\partial X_c^2} \cdot Q_1(A(X_c, P)) + \frac{\partial P}{\partial X_c} \cdot \left(\frac{\partial Q_1}{\partial X_c}\right)\right) + \frac{A_1(P)}{\partial P} \cdot \frac{\partial P}{\partial T} = 0 \tag{2.6}$$

yielding the relevant viscous-elastic time scale $t_f^*$,

$$t_f^* = \frac{\sigma_r \mu}{p^* \varepsilon_1^2} = \frac{a_1^* \mu}{a_0^* p^* \varepsilon_1^2}. \tag{2.7}$$

The resulting convection-diffusion equation has a material non-linearity due to the pressure dependent coefficients $Q_1(A(X_c, P))$ and $A_1(P)/\partial P$ which govern the diffusivity coefficient. Setting both to constant, degenerates (2.6), to become a linear diffusion equation losing both its non-linearity and its transport term,

$$-\left(\frac{\partial^2 P}{\partial X_c^2} \cdot Q_1\right) + \frac{\partial P}{\partial T} = 0 \tag{2.8}$$

and the characteristic time scale $t_f^*$, becomes,

$$t_f^* = \frac{\mu}{a_0^* \varepsilon_1^2} \left(\frac{\partial a_1(p)}{\partial p}\right)\bigg|_{p=p_0}. \tag{2.9}$$

Substituting normalized parameters into (1.3) yields,

$$\frac{\partial^2}{\partial z_s^2}\left[\left(\frac{\partial^2 D_s}{\partial z_s^2} + \Phi(Z_s) \cdot P(X_c, T)\frac{\partial \psi}{\partial (p/E)}\right)\right] = -\frac{\partial^2 D_s}{\partial T^2} \cdot \left(\frac{t_s^*}{t_f^*}\right)^2 + F(Z_s, T) \tag{2.10}$$

and order of magnitude analysis yields

$$d_s^* = \frac{l_s^2 \cdot \phi^* \cdot \mu \cdot u^*}{E \cdot \varepsilon_1^2 \cdot l}, \tag{2.11}$$

And

$$f^* = \frac{\phi^* p^* I}{l_s^2} \tag{2.12}$$

where $t_s^* = \sqrt{\rho_s l_s^4 / EI}$. The ratio $t_s^*/t_f^*$ determines the relevant regime for beam dynamics. For $t_s^*/t_f^* \gg 1$, the viscous-elastic interaction propagates significantly faster than beam response, and the beam behaves as responding to a spatially uniform, time varying pressure. This case was thoroughly investigated in previous work[12]. For $t_s^*/t_f^* = O(1)$, the spatial pressure variation within the network must be accounted in the analysis and solid inertia will take part in the interaction. For $t_s^*/t_f^* \ll 1$, beam inertial term is negligible and the dynamics of the beam reflects variations in the internal pressure field.

Rewriting (2.10) to solve for $P(Z_s, T)$ we obtain an integro-differential equation correlating fluid pressure to beam deflection,

$$P(Z_s, T) = \left(\Phi(Z_s)\frac{\partial \psi}{\partial (p/E)}\right)^{-1}\left(\int_0^{Z_s}\int_0^{\zeta}\left(F(\eta, T) - \frac{\partial^2 D_s}{\partial T^2}\left(\frac{t_s^*}{t_f^*}\right)^2\right)d\eta d\zeta - \frac{\partial^2 D_s}{\partial Z_s^2}\right). \tag{2.13}$$

Applying coordinate transformation (1.5) to (2.8) and substituting (2.13), we obtain the general form of the governing partial, nonlinear fourth-order integro-differential (see Appendix A). Setting $\partial \psi / \partial (p/E)$, $\partial A_1(P)/\partial P$, $Q_1(A(Z_s, P))$ and $\Phi(Z_s)$ to be constants, we define all material properties and channel distribution to be spatially uniform, and obtain a simplified governing equation of the form,

$$\left(\frac{ls}{l} + \Phi\right)^2 \frac{\partial A_1}{\partial P}\left(\int_0^{Z_s}\int_0^{\zeta}\left(\frac{\partial F(\eta, T)}{\partial T} - \frac{\partial^3 D_s}{\partial T^3}\left(\frac{t_s^*}{t_f^*}\right)^2\right)d\eta d\zeta - \frac{\partial^3 D_s}{\partial Z_s^2 \partial T}\right)$$
$$- Q_1\left(F(Z_s, T) - \frac{\partial^2 D_s}{\partial T^2}\left(\frac{t_s^*}{t_f^*}\right)^2 - \frac{\partial^4 D_s}{\partial Z_s^4}\right) = 0. \tag{2.14}$$

The parameter $\partial \psi / \partial (p/E)$, which scales our deflection, now appears only in the boundary and initial conditions.

Equation (2.14) requires four boundary conditions and three initial conditions, as is required to satisfy (2.6) and (2.10). First, we define four boundary conditions over $D_s$, these are formulated as a set of geometric and dynamic conditions. While geometric conditions, over deflection,

$$D_s(Z_s)|_{(\alpha,T)} = G_1(Z_s) \tag{2.15}$$

or slope,

$$\left(\frac{\partial D_s}{\partial T}\right)\bigg|_{(\alpha,T)} = G_2(Z_s), \tag{2.16}$$

remain unchanged compared with standard Euler-Bernoulli boundary conditions, dynamic conditions, such as moment $M = G_3(T)$ and shear force $V = G_4(T)$, act only on $D_e$, as they represent conditions applied on the total elastic energy stored within the beam material $EI\,(\partial^2 d_e)/(\partial z_s^2)$ as such require special attention. As shown in previous work[12], we use a correlation to relate pressure-to-change-in-slope,

$$\frac{\partial^2 d_n}{\partial x^2} = -\phi \frac{p}{E} \frac{\partial \psi}{\partial (p/E)}\left(\frac{p}{E} = 0, \nu, \frac{z_i}{h}, \frac{d_i}{h}\right). \tag{2.17}$$

Adjusting to current nomenclature and substituting the normalized parameters, equation (2.17) now reads,

$$\frac{\partial^2 D_C}{\partial Z_s^2} = -\Phi(Z_s) \cdot P(Z_s, T) \frac{\partial \psi}{\partial (p/E)}. \tag{2.18}$$

Recalling $d_s = d_e + d_c$, we can now formulate dynamic boundary conditions as,

$$\left(\frac{\partial^2 D_s}{\partial Z_s^2}\right)\bigg|_{(\alpha,T)} = \frac{G_3(T)}{EI} - \Phi(Z_s) \cdot P(\alpha, T) \frac{\partial \psi}{\partial (p/E)} \tag{2.19}$$

and

$$\left(\frac{\partial^3 D_s}{\partial Z_s^3}\right)\bigg|_{(\alpha,T)} = \frac{G_4(T)}{EI} - \frac{\partial}{\partial Z_s}\left(\Phi(Z_s) \cdot P(\alpha, T) \frac{\partial \psi}{\partial (p/E)}\right), \tag{2.20}$$

where $P(\alpha, T)$ is the known pressure set at $\alpha = 0$ or $\alpha = 1$. The initial conditions are

$$D_s(Z_s, 0) = G_5(Z_s), \tag{2.21}$$

$$\left(\frac{\partial D_s}{\partial T}\right)\bigg|_{(Z_s,0)} = G_6(Z_s). \tag{2.22}$$

and

$$P(Z_s, 0) = G_7(Z_s). \tag{2.23}$$

## 3. Results

In Figs. 2-5 we illustrate the deflection and pressure field of a beam with an embedded fluidic network due to viscous-elastic dynamics. In all presented cases the beam and channel network are of identical geometry and physical properties. We set the time scale ratio $t_s^*/t_f^* \ll 1$, thus solid inertia is negligible and beam dynamics reflect the variation in internal pressure field over time and space. Our material and structure parameters are set to constants, $\partial A_1(P)/\partial P = 1$, $Q_1(A(Z_s, P)) = 0.3$, $\partial \psi/\partial (p/E) = 1$, and $\Phi(Z_s) = 1$. All figures present solutions of (2.14).

Fig. 2 illustrates a basic case of a suddenly applied inlet pressure, propagating via diffusion down the length of the channel, creating deflection of the beam. The configuration is of a cantilever beam, clamped at $Z_s = 0$ and free at $Z_s = 1$, starting from rest $D_s(Z_s, T \leq 0) = 0$. A ramp pressure boundary condition $P(Z_s = 1, T) = H(T)$ is introduced, where $H(T)$ is the Heaviside function, and the channel is sealed at $Z_s = 0$, $\partial P/\partial Z_s (Z_S = 0) = 0$. Beam deflection $D_s$ (part a), scaled beam deflection $D_s/D_s(1,T)$ (part b) and fluid pressure $P$ (part c) are presented vs. $Z_s$ for various times, $T = 0.05, 0.16, 0.3, 0.7, 1.6, 10$. Both the pressure and the deformation fields propagate into the beam with speed of order of magnitude of $O(1)$ (in dimensional terms $O(l/t_f^*)$), and the pressure at the free-end $Z_s = 0$ starts increasing only after $T \approx 0.7$.

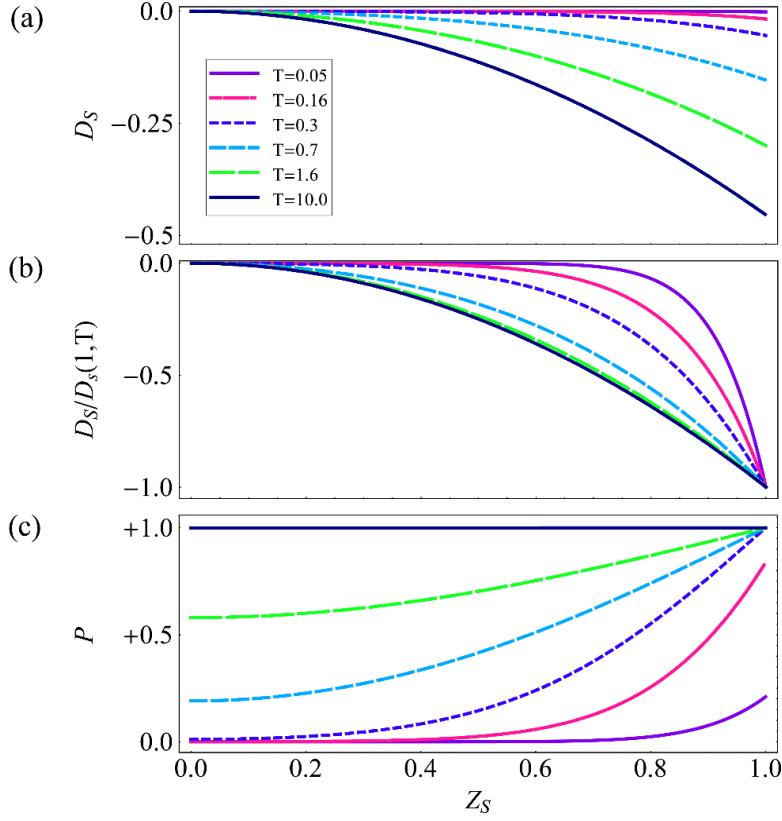

FIG. 2. *Cantilever beam clamped at $Z_s = 0$, starting from rest $D_s(Z_s, T \leq 0) = 0$, with boundary conditions of ramp pressure $P(Z_s = 1, T) = H(T)$, where $H(T)$ is the Heaviside function, and $\partial P/\partial Z_s (Z_S = 0) = 0$. Beam deformation (a) Beam deformation normalized by $D_s(1,T)$ (b) and pressure (c) vs. $Z_S$ for several normalized times.*

Introducing oscillating pressure with significantly smaller oscillation period compared to the viscous elastic time scale, we can expect only part of the beam to deflect. In Fig.3 we examine a cantilever beam where the inlet pressure at $Z_s = 1$ is $P(1,T) = P_1 sin(2\pi \mathcal{F}_p \cdot T)$, whereas all other boundary conditions are identical to the beam in Fig. 2. Using varying inlet frequency $\mathcal{F}_p$, matched with pressure amplitude $P_1$, we are able to control the length of the beam engaged in the oscillating motion. Setting $\mathcal{F}_p = 0.1$ and $P_1 = 0.02$ we are able to produce beam deflection of $D_s \approx 0.006$ at $Z_s = 1$, with beam length engaged in the deflection starting from $Z_s \geq 0.1$, marked in solid blue. Setting $\mathcal{F}_p = 10$ and $P_1 = 1.5$ we are able to produce an identical beam deflection at tip, $Z_s = 1$, with beam length engaged in deflection starting from $Z_s \geq 0.6$, marked in solid red.

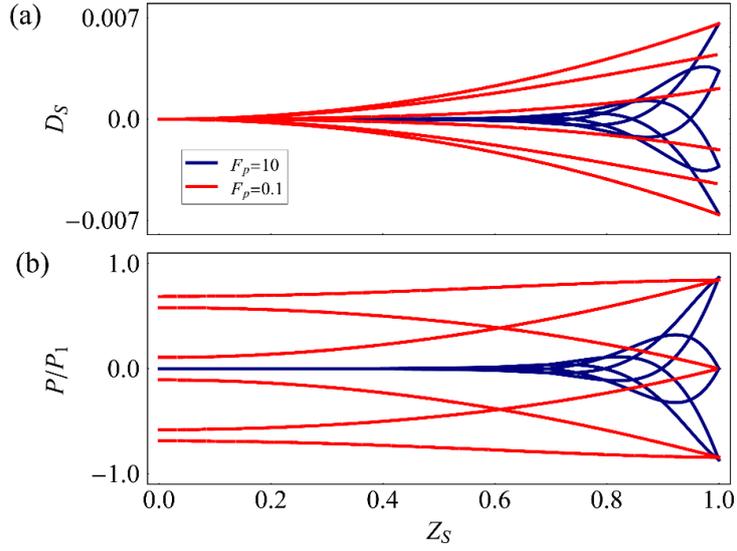

*FIG. 3. (a) Cantilever beam starting from rest with an oscillating pressure inlet introduced at $Z_s = 1$ as $P(1,T) = P_1 sin(2\pi \mathcal{F}_p \cdot T)$ Setting $\mathcal{F}_p = 0.1$ and $P_1 = 0.02$ (solid red, Right scale) and $\mathcal{F}_p = 10$ and $P_1 = 1.3$ (solid blue, left scale), deflection is illustrated via centerline and each cycle is divided into six equal parts. (b) Respective Pressure profiles along beam length.*

In Figs. 4 and 5 we illustrate using oscillating pressure introduced to inlets at both ends, as a mechanism to create wave-like deformation fields. Matching frequency, amplitude, and phase angle, we are able to reproduce an inertia-like standing and moving waves, in a system void of inertia and transport terms, via viscous effects alone.

In Fig.4 we illustrate the deflection $D_s$ (a) and pressure $P$ (b) vs. $Z_s$ fpr a simply supported beam, hinged at $Z_s = 0$ and at $Z_s = 1$. Pressure is applied at both ends, given by $P(0,T) = 1.5 \cdot sin(2\pi \cdot 0.2 \cdot T + \pi/2)$ and $P(1,T) = 1.5 \cdot sin(2\pi \cdot 0.2 \cdot T + 3\pi/2)$. The obtained viscous-elastic deformation field (solid blue lines) closely follow an inertial standing wave of the form $D_s(Z_s, T) = 0.01(sin(2\pi Z_s + 2\pi \cdot 0.2 \cdot T) + sin(2\pi Z_s - 2\pi \cdot 0.2 \cdot T))$, (dashed black lines).

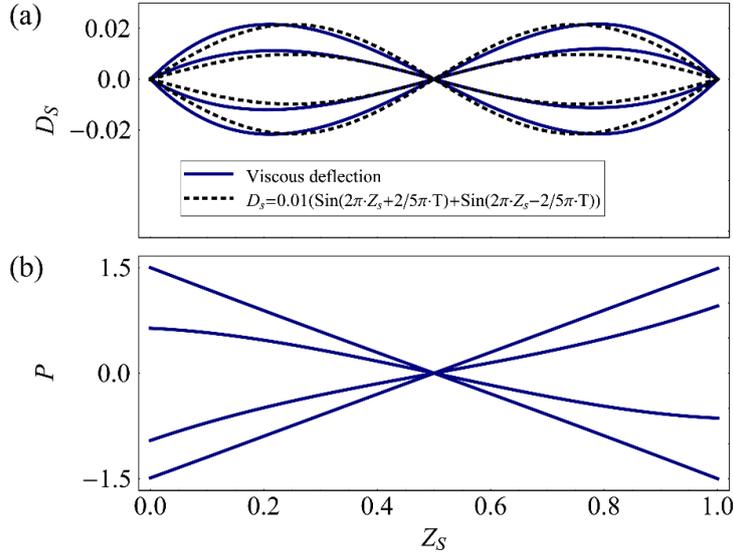

FIG. 4. *Simply supported beam with an oscillating pressure introduced to inlets at both ends, $P(0,T) = 1.5 \cdot sin(2\pi \cdot 0.2 \cdot T + \pi/2)$ and $P(1,T) = 1.5 \cdot sin(2\pi \cdot 0.2 \cdot T + 3\pi/2)$. Viscous deflection (Solid blue) and an inertial standing wave, $D_s(Z_s, T) = 0.01(sin(2\pi Z_s + 2\pi \cdot 0.2 \cdot T) + sin(2\pi Z_s - 2\pi \cdot 0.2 \cdot T))$, (dashed black) are compared. (b) Respective pressure profiles along beam length.*

In Fig. 5 our beam support at $Z_s = 0$ is set to oscillate as $(D_s)|_{(0,T)} = -0.0375 \sin(2\pi \cdot 0.5 \cdot T)$ and $(\partial D_s/\partial Z_s)|_{(0,T)} = 2\pi \cdot 0.5 \cdot 0.0375 \cdot \cos(2\pi \cdot 0.5 \cdot T)$. The beam is free at $Z_s = 1$. Pressure is introduced to inlets at both ends $P(0,T) = 0.7 \cdot sin(2\pi \cdot 0.5 \cdot T + \pi)$ and $P(1,T) = 0.8 \cdot sin(2\pi \cdot 0.5 \cdot T + \pi/2)$. In this case the deformation of the beam (blue solid line) closely match a moving wave given by $D_s(Z_s, T) = 0.0375 \cdot \sin(2\pi \cdot 0.5 \cdot Z_s - 2\pi \cdot 0.5 \cdot T)$ (dashed black lines).

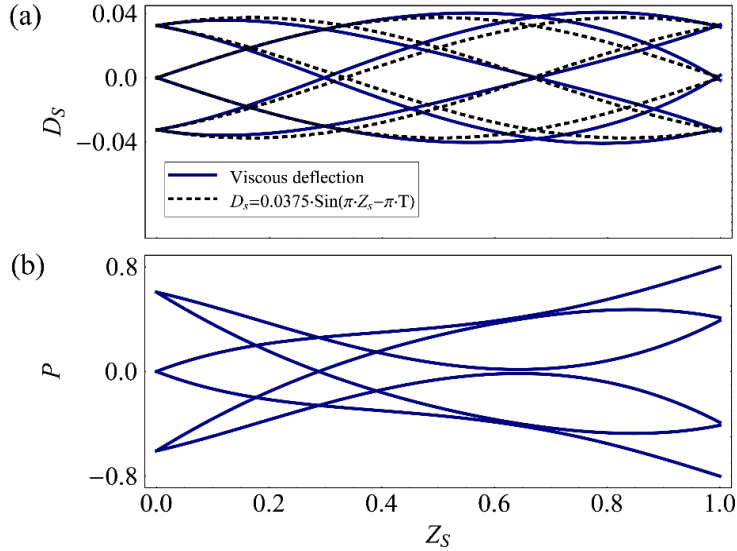

FIG. 5. *Beam externally oscillated at $Z_s = 0$, $(D_s)|_{(0,T)} = -0.0375 \sin(2\pi \cdot 0.5 \cdot T)$ and $(\partial D_s/\partial Z_s)|_{(0,T)} = 2\pi \cdot 0.5 \cdot 0.0375 \cdot \cos(2\pi \cdot 0.5 \cdot T)$, free at $Z_s = 1$. Oscillating pressure is introduced to inlets at both ends $P(0,T) = 0.7 \cdot sin(2\pi \cdot 0.5 \cdot T + \pi)$ and $P(1,T) = 0.8 \cdot sin(2\pi \cdot 0.5 \cdot T + \pi/2)$. (a) Deformation due to viscous-elastic dynamics (solid blue lines) and an inertial moving wave, $D_s(Z_s, T) = 0.0375 \cdot \sin(2\pi \cdot 0.5 \cdot Z_s - 2\pi \cdot 0.5 \cdot T)$, (dashed black lines).. (b) Respective Pressure profiles along beam length.*

# 5. Appendix A

## 5.1. General formulation

Applying coordinate transformation (1.5) to (2.8) and substituting (2.13), we obtain the general governing design equation as a fourth-order, nonlinear, partial, integro-differential equation,

$$\left(\frac{\text{ls}}{l} + \Phi(Z_s)\right)^3 \frac{\partial A_1(P)}{\partial P}\left(\left(\Phi(Z_s) \cdot \frac{\partial \psi}{\partial (p/E)}\right)^{-1}\left(\iint_{0\;0}^{Z_s\;\zeta}\left(\frac{\partial F(\eta,T)}{\partial T} - \frac{\partial^3 D_s}{\partial T^3} \cdot \frac{t_s^{*2}}{t_f^{*2}}\right) d\eta d\zeta - \frac{\partial^3 D_s}{\partial Z_s^2 \partial T}\right)\right) - \left(\frac{\text{ls}}{l} + \Phi(Z_s)\right)\left(\frac{\partial Q_1}{\partial Z_s}\left(\frac{\partial}{\partial Z_s}\left(\Phi(Z_s) \cdot \frac{\partial \psi}{\partial (p/E)}\right)^{-1}\left(\iint_{0\;0}^{Z_s\;\zeta}\left(F(\eta,T) - \frac{\partial^2 D_s}{\partial T^2} \cdot \frac{t_s^{*2}}{t_f^{*2}}\right) d\eta d\zeta - \frac{\partial^2 D_s}{\partial Z_s^2}\right) + \left(\Phi(Z_s) \cdot \frac{\partial \psi}{\partial (p/E)}\right)^{-1}\left(\int_0^{Z_s}\left(F(\eta,T) - \frac{\partial^2 D_s}{\partial T^2} \cdot \frac{t_s^{*2}}{t_f^{*2}}\right) d\eta - \frac{\partial^3 D_s}{\partial Z_s^3}\right)\right) + Q_1(A(Z_s,P))\left(\frac{\partial^2}{\partial Z_s^2}\left(\Phi(Z_s) \cdot \frac{\partial \psi}{\partial (p/E)}\right)^{-1}\left(\iint_{0\;0}^{Z_s\;\zeta}\left(F(\eta,T) - \frac{\partial^2 D_s}{\partial T^2} \cdot \frac{t_s^{*2}}{t_f^{*2}}\right) d\eta d\zeta - \frac{\partial^2 D_s}{\partial Z_s^2}\right) + 2 \cdot \frac{\partial}{\partial Z_s}\left(\Phi(Z_s) \cdot \frac{\partial \psi}{\partial (p/E)}\right)^{-1}\left(\int_0^{Z_s}\left(F(\eta,T) - \frac{\partial^2 D_s}{\partial T^2} \cdot \frac{t_s^{*2}}{t_f^{*2}}\right) d\eta - \frac{\partial^3 D_s}{\partial Z_s^3}\right) + \left(\Phi(Z_s) \cdot \frac{\partial \psi}{\partial (p/E)}\right)^{-1}\left(\left(F(Z_s,T) - \frac{\partial^2 D_s}{\partial T^2} \cdot \frac{t_s^{*2}}{t_f^{*2}}\right) - \frac{\partial^4 D_s}{\partial Z_s^4}\right)\right) + Q_1(A(Z_s,P))\Phi'(Z_s)\left(\frac{\partial}{\partial Z_s}\left(\Phi(Z_s) \cdot \frac{\partial \psi}{\partial (p/E)}\right)^{-1}\left(\iint_{0\;0}^{Z_s\;\zeta}\left(F(\eta,T) - \frac{\partial^2 D_s}{\partial T^2} \cdot \frac{t_s^{*2}}{t_f^{*2}}\right) d\eta d\zeta - \frac{\partial^2 D_s}{\partial Z_s^2}\right) + \left(\Phi(Z_s) \cdot \frac{\partial \psi}{\partial (p/E)}\right)^{-1}\left(\int_0^{Z_s}\left(F(\eta,T) - \frac{\partial^2 D_s}{\partial T^2} \cdot \frac{t_s^{*2}}{t_f^{*2}}\right) d\eta - \frac{\partial^3 D_s}{\partial Z_s^3}\right)\right) = 0 \quad (5.1)$$